# FIRST STEPS TOWARD INCORPORATING IMAGE BASED DIAGNOSTICS INTO PARTICLE ACCELERATOR CONTROL SYSTEMS USING CONVOLUTIONAL NEURAL NETWORKS

A.L. Edelen[†], S.G. Biedron[1,2], S.V. Milton[2], Colorado State University, Fort Collins, CO
J.P. Edelen, Fermilab[*], Batavia, IL
[1]also at University of Ljubljana, Ljubljana, Slovenia [2]also at ElementAero, Chicago, IL

*Abstract*

At present, a variety of image-based diagnostics are used in particle accelerator systems. Often times, these are viewed by a human operator who then makes appropriate adjustments to the machine. Given recent advances in using convolutional neural networks (CNNs) for image processing, it should be possible to use image diagnostics directly in control routines (NN-based or otherwise). This is especially appealing for non-intercepting diagnostics that could run continuously during beam operation. Here, we show results of a first step toward implementing such a controller: our trained CNN can predict multiple simulated downstream beam parameters at the Fermilab Accelerator Science and Technology (FAST) facility's low energy beamline using simulated virtual cathode laser images, gun phases, and solenoid strengths.

## INTRODUCTION

Recently, convolutional NNs (CNNs) have yielded impressive results in the area of computer vision, especially for image recognition tasks [1]. They are also starting to be used in physics-related applications, such as automatic classification of galaxies [2] and neutrino events [3]. Given the present success of CNNs, it may now be possible to use them as a means of incorporating image diagnostics directly into particle accelerator control systems. While this could be done in a variety of ways (e.g. even just using traditional control methods with extracted image information), the avenue we have chosen to pursue is to create a NN controller that directly processes image data using some initial convolutional layers.

In support of this, we are investigating automated control over the photocathode electron gun at the Fermilab Accelerator Science and Technology (FAST) facility [4,5], specifically focused on automated beam alignment and tuning of the solenoid strength and gun phase. For a given laser system it is not always easy to produce a top-hat transverse laser profile, and any asymmetries in the initial laser distribution can impact the electron beam parameters. As such, it could be useful to train a controller to take a measured laser distribution image (here, the virtual cathode image) and yield optimal gun phase and solenoid strength settings (as determined by downstream beam parameters). In principle, if one had fine control over the transverse laser distribution itself, one could also include it as a controllable parameter.

To develop this type of controller, the first step is ensuring that a NN can adequately predict the beam parameters from various input distribution images, the gun phase settings, and the solenoid strengths. This process model can then be used to help train a NN controller. To this end, we have created a NN model that uses physics-based simulations of FAST as training data. It is also worth noting that by using simulation data to train the NN, we've created a fast-executing representation of the dynamics that could also be used in model predictive control, offline optimization studies, or quick tests in the control room without perturbing the machine. In this paper, we discuss our simulation studies, provide an overview of the NN architecture, and show the NN's performance in predicting beam parameters at the exit of the gun and the second capture cavity.

## PARMELA SIMULATIONS

Simulations of the first 8 meters of the FAST low energy beamline were conducted using PARMELA [6]. Included in the simulations are the electron gun, both superconducting capture cavities, and the intermediate beam-line elements. The locations of the cavities and beam-line elements were taken from a Fall 2015 mechanical survey. Two sets of simulation scans were conducted: one set of fine scans to predict beam parameters after the gun, and one set of coarse scans to predict beam parameters after the second capture cavity (CC2). The gun phase was scanned from -180° to 180° in 10° and 5° steps, and the solenoid strength was scanned from 0.5 to 1.5 in 10% and 5% steps, where 1.0 represents the nominal setting that produces the peak axial field of 1.8 kGauss. Prior to scaling of the field maps, the bucking coil was tuned to produce zero magnetic field on the cathode. The field maps of the solenoid assembly, gun, and capture cavities used for the PARMELA simulations were generated using Poisson Superfish [7].

For the gun studies, we used three initial top-hat beam distributions with different radii. For the simulations up through CC2, we used nine different beam distributions. Three of these were the same top-hat distributions used for the gun simulations, and the other six were Gaussian transverse distributions with different RMS widths in x and y. These distributions were also converted into the simulated virtual cathode images (e.g. Figure 1) that were later used in training the NN. Prior simulation results using initial beam distributions derived from measured virtual cathode images suggest this is a sound approach. For all cases, the longitudinal distribution of the beam was Gaussian with a bunch length of 3 ps.

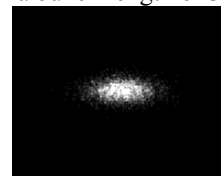

Figure 1: Example of an initial beam distribution image used as input to the NN.

---

Figure 2 shows some representative data from the simulations. For the CC2 data, one can see asymmetry in the transverse emittance caused by the initial beam distribution. Additionally, all of the curves are nonlinear with respect to both solenoid strength and phase. To give a sense of scale of the target data used in training, Table 1 shows the range of all predicted parameters.

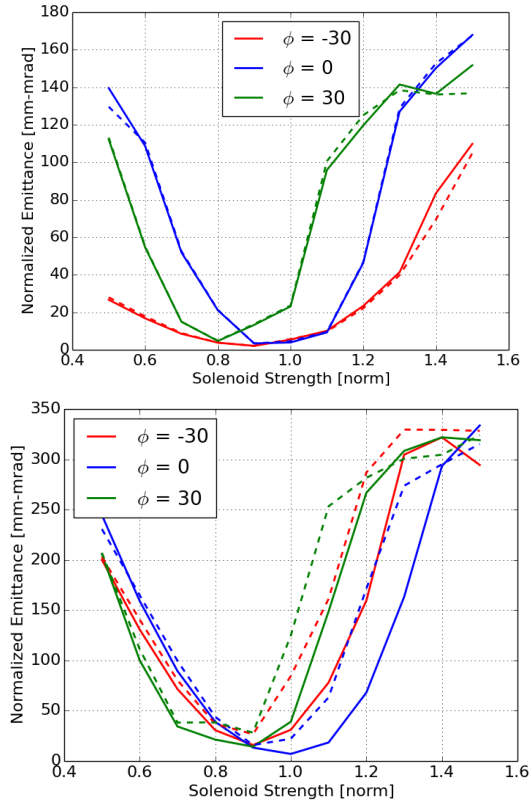

Figure 2: These plots show the transverse emittances as a function of the normalized solenoid strength for three gun phases. Dashed lines denote x emittance, and solid lines denote y emittance. The top plot is for a top-hat initial beam distribution with a width of 0.6 mm. The lower plot is for a Gaussian initial beam distribution with an x RMS width of 0.6 mm and a y RMS width of 1.2 mm.

Table 1: Max and Min Values for Predicted Parameters

| Param. | Max Gun | Min Gun | Max CC2 | Min CC2 |
|---|---|---|---|---|
| $N_p$ | 5001 | 1015 | 5001 | 1004 |
| $\varepsilon_{nx}$ [m-rad] | 2.5e-4 | 1.6e-6 | 4.0e-4 | 9.1e-7 |
| $\varepsilon_{ny}$ [m-rad] | 2.4e-4 | 1.6e-6 | 4.0e-4 | 8.5e-7 |
| $\alpha_x$ [rad] | 14.1 | -775.1 | 0.8 | -149.8 |
| $\alpha_y$ [rad] | 14.5 | -797.0 | 0.7 | -154.5 |
| $\beta_x$ [m/rad] | 950.4 | 7.9e-2 | 820.2 | 0.7 |
| $\beta_y$ [m/rad] | 896.8 | 8.4e-2 | 845.7 | 0.81 |
| E [MeV] | 4.6 | 3.2 | 47.2 | 42.8 |

## NEURAL NETWORK MODELS

Instead of a classification task (as is most common), here we use a CNN in a regression task (see Figure 3). We also adopt a hybrid structure that joins a CNN and fully-connected NN to incorporate both image-based data and non-image-based data into the model. The network uses the simulated virtual cathode image, the solenoid strength, and the gun phase as inputs. The latter two inputs bypass the convolutional layers of the network and are later combined in a final fully-connected set of layers. The outputs are the number of transmitted particles ($N_p$), the transverse emittances ($\varepsilon_{nx}$, $\varepsilon_{ny}$), the average beam energy (E), and the transverse alpha and beta function values ($\alpha_x$, $\alpha_y$, $\beta_x$, $\beta_y$).

A variety of NN structures were examined (including various number of layers, number of filters per layer, filter sizes, activation functions, etc.). The chosen network consists of 3 convolutional layers: 16 5x5 filters, followed by 16 3x3 filters, followed by 10 3x3 filters. These are followed by 3 fully-connected layers with 150, 70, and 8 neurons respectively. With the exception of the linear output layer, hyperbolic tangent activation functions are used.

The model was trained using a combination of the ADADELTA [8] and Adam [9] optimization algorithms, and the network weights were initialized using the layer-by-layer method described in [10] with a uniform distribution. The network biases were initialized using a normal distribution with standard deviation of 0.01 and a mean of 0. For the gun data, the training set consisted of 1395 data points and the validation set consisted of 200 data points. For the CC2 data, the training set consisted of 894 data points and the validation set consisted of 600 data points. The CC2 data consists of more images and coarser scans for each image, hence the different ratio of training to validation data. Note that because we were data-limited in this case, we used randomly sampled validation data from across the data set and did not have a test set. For further testing, new distribution images should be used to ensure that the NN can interpolate between them sufficiently well. After training the gun network, we used its weights as an initial solution to begin training the CC2 network.

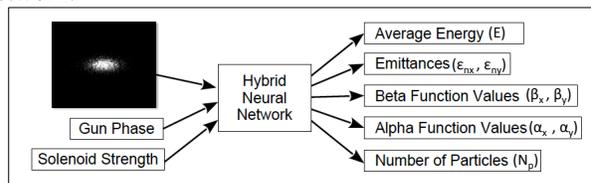

Figure 3: Neural network inputs and outputs.

## MODEL PERFORMANCE

Tables 2 and 3 show the NN's performance in terms of mean absolute error (MAE) and standard deviation (STD) over the training and validation sets. Table 2 shows the model performance for the data after the gun, and Table 3 shows the performance for the data after CC2. For the gun, all MAEs are between 0.4% and 1.8% of the parame-ter ranges, and for CC2, all MAEs are between 0.9% and 3.1% of the parameter ranges.

In Figure 4, we highlight two representative data sets to show the NN's performance in predicting downstream beam parameters.

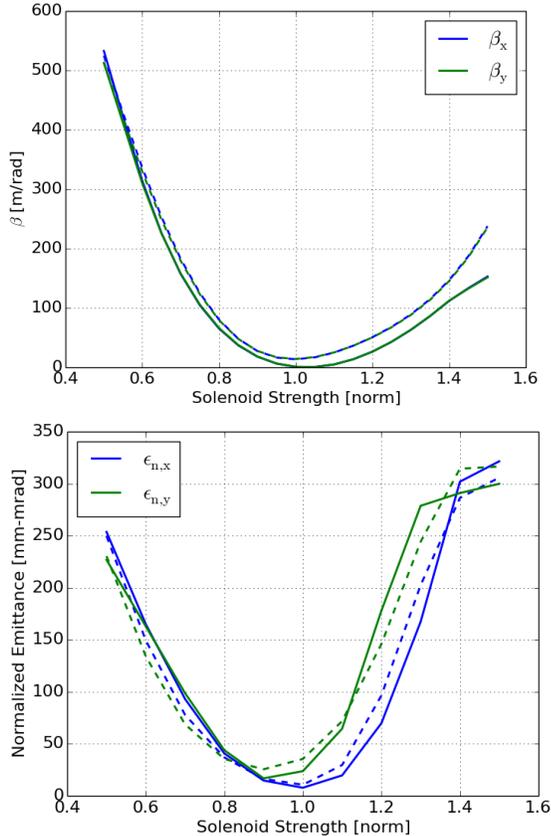

Figure 4: The top plot shows transverse beta function values after the gun as a function of solenoid strength for a top-hat initial beam distribution. The bottom plot shows normalized transverse emittances after CC2 as a function of solenoid strength for an asymmetric Gaussian beam distribution (the one shown in Figure 1). Both are for an RF phase of 0°. The dashed lines are NN predictions and the solid lines are simulated values.

Table 2: Model Performance at Gun Exit

| Param. | Train. MAE | Train. STD | Val. MAE | Val. STD |
|---|---|---|---|---|
| $N_p$ | 69.5 | 79.8 | 70.7 | 75.7 |
| $\varepsilon_{nx}$ | 2.3e-6 | 3.5e-6 | 2.4e-6 | 3.2e-6 |
| $\varepsilon_{ny}$ | 2.3e-6 | 3.4e-6 | 2.4e-6 | 3.2e-6 |
| $\alpha_x$ | 9.0 | 14.9 | 10.9 | 16.0 |
| $\alpha_y$ | 8.8 | 15.3 | 10.8 | 16.1 |
| $\beta_x$ | 12.1 | 17.6 | 14.8 | 18.9 |
| $\beta_y$ | 11.7 | 16.7 | 14.3 | 17.9 |
| E | 4.9e-3 | 4.9e-3 | 5.5e-3 | 6.0e-3 |

Table 3: Model Performance at CC2 Exit

| Param. | Train. MAE | Train. STD | Val. MAE | Val. STD |
|---|---|---|---|---|
| $N_p$ | 103.7 | 141.2 | 123.3 | 176.8 |
| $\varepsilon_{nx}$ | 1.0e-5 | 1.2e-5 | 1.2e-5 | 1.6e-5 |
| $\varepsilon_{ny}$ | 1.0e-5 | 1.3e-5 | 1.2e-5 | 1.5e-5 |
| $\alpha_x$ | 3.4 | 6.6 | 3.1 | 5.9 |
| $\alpha_y$ | 3.4 | 6.6 | 3.1 | 5.9 |
| $\beta_x$ | 16.3 | 33.5 | 14.7 | 27.8 |
| $\beta_y$ | 16.4 | 33.6 | 14.8 | 27.5 |
| E | 4.0e-2 | 3.9e-2 | 4.6e-2 | 6.2e-2 |

## CONCLUSION AND NEXT STEPS

We have shown that our convolutional/fully-connected NN is capable of predicting simulated downstream beam parameters given solenoid strengths, gun phases, and simulated virtual cathode images. This is an important first step toward creating an NN controller that can directly use image diagnostics. Despite the small number of samples in the training set and the large number of predicted parameters, the network performs fairly well on average. All mean absolute errors are between 0.4% and 3.1% of the parameter ranges, and this is likely to improve with additional training data. In addition, this fast-executing model could already be used on its own for rapid optimization studies.

Presently, we are extending this work to include measured training data from the machine. For those studies, beam alignment will be used as an additional predicted parameter. Once the model is updated with measured data, we plan to train a neural network controller and test it on the machine.

## ACKNOWLEDGMENT

The authors thank Curt Baffes and the Fermilab mechanical group for providing the machine survey data. They also thank Philippe Piot for discussions on simulating the FAST LINAC, providing prior simulations, and for the help with the Superfish models of the cavities and the solenoid. They are also grateful to the Fermilab Computing Division, whose high performance computing resources were essential for training the neural network.